# 数据要素对缩小中国城乡消费差距的影响：机制与政策分析

马铭璞

（天津大学管理与经济学部，天津 300072）

## 摘 要

城乡居民消费差距作为社会发展中重要的指标之一，直接反映了城乡经济社会发展的不平衡性。数据要素作为新质生产力的重要组成部分，在信息时代的推动下，对经济发展和人民生活水平提升具有重要意义。本研究通过固定效应回归模型、系统 GMM 回归模型和中介效应模型的分析，发现数据要素的发展水平在一定程度上促进了城乡消费差距的缩小，同时城乡收入差距这一中介变量在数据要素与消费差距之间发挥着重要作用，中介效应显著。研究结果表明，数据要素的进步能够通过减小城乡收入差距来促进城乡居民消费水平的均衡，为实现共同富裕和促进城乡协调发展提供了理论支持和政策建议。

在此基础上，本文强调了数据要素发展与城乡居民消费差距之间复杂的关联性，提出了推动数据要素市场发展、加强数字经济与电子商务建设、促进城乡融合发展等政策建议。综合来看，数据要素的发展既是城乡消费差距缩小的重要路径，也是推动中国经济社会平衡发展的关键之一。这一研究对理解城乡消费差距形成机制，完善城乡经济发展政策具有一定的理论与实践意义。

**关键词**：城乡消费差距，数据要素发展，中介效应，固定效应回归，GMM 回归

# The Impact of Data Elements on Narrowing the Urban-Rural Consumption Gap in China: Mechanisms and Policy Analysis

Mingpu Ma*

* College of Management and Economics, Tianjin University, Tianjin 300072, China
E-mail: m125_0409@tju.edu.cn

## ABSTRACT

The urban-rural consumption gap, as one of the important indicators in social development, directly reflects the imbalance in urban and rural economic and social development. Data elements, as an important component of New Quality Productivity, are of significant importance in promoting economic development and improving people's living standards in the information age. This study, through the analysis of fixed-effects regression models, system GMM regression models, and the intermediate effect model, found that the development level of data elements to some extent promotes the narrowing of the urban-rural consumption gap. At the same time, the intermediate variable of urban-rural income gap plays an important role between data elements and consumption gap, with a significant intermediate effect. The results of the study indicate that the advancement of data elements can promote the balance of urban and rural residents' consumption levels by reducing the urban-rural income gap, providing theoretical support and policy recommendations for achieving common prosperity and promoting coordinated urban-rural development.

Building upon this, this paper emphasizes the complex correlation between the development of data elements and the urban-rural consumption gap, and puts forward policy suggestions such as promoting the development of the data element market, strengthening the construction of the digital economy and e-commerce, and promoting integrated urban-rural development. Overall, the development of data elements is not only an important path to reducing the urban-rural consumption gap but also one of the key drivers for promoting the balanced development of China's economic and social development. This study has a certain theoretical and practical significance for understanding the mechanism of the urban-rural consumption gap and improving policies for urban-rural economic development.

# 目 录

# 一、引言

城乡居民消费差距过大作为城乡发展不平衡的重要表征，一直是中国社会发展面临的突出问题。相比城镇居民，中国农村居民消费增长的广度与深度仍然不足，城乡居民消费差距远远高于国际平均水平。要想实现我国居民消费水平的进一步提高，缩小城乡消费差距迫在眉睫。如何缩小城乡消费差距、破解城乡鸿沟，成为共同富裕政策锚定的重要发力点。

随着信息时代的来临，我国大数据等科学技术的不断进步正推动着社会生产力与经济的不断革新与发展。数据要素作为大数据处理与应用的基础，其所释放的动能在经济发展、社会服务等方面都发挥着重要的意义与价值。一方面，由中国特殊的城乡二元结构导致城乡数字鸿沟问题日益尖锐，城乡居民消费差距有进一步扩大的可能；另一方面，数据要素发展显著降低了跨区域的信息不对称程度，突破了城乡之间的物理阻碍，肩负起助力农业发展、农村繁荣和农民致富的时代使命。因此，在扎实推进共同富裕进程中迫切需要探索数据要素的发展举措，促进城乡协调发展。

当前，大力推进现代化产业升级、加快发展新质生产力成为我国政府的重要目标。新质生产力是指创新起主导作用，摆脱传统经济增长方式、生产力发展路径，具有高科技、高效能、高质量特征，符合新发展理念的先进生产力质态。习近平指出，新质生产力以全要素生产率大幅提升为标志。而数据要素正是全要素生产率的重要组成部分，在经济社会发展中起着重要作用。

基于数据要素的重要作用，我国也出台了一系列关于促进数据要素发展的政策文件。中共中央于 2020 年 3 月 30 日颁布的《关于构建更加完善的要素市场化配置体制机制的意见》中首次从国家层面提出数据要素的概念，明确将数据要素列入土地、劳动力、资本、技术等四大生产要素中；2022 年 12 月 9 日，中共中央、国务院印发《关于构建数据基础制度更好发挥数据要素作用的意见》初步搭建了我国数据的基础制度体系，为数据要素发展提供了更有效的保障；2023 年 12 月 31 日，国家数据局同其他相关部门制定了《"数据要素×"三年行动计划（2024—2026）》，再次强调了数据要素是推动社会经济高质量发展的关键，对数

据要素应如何更好的促进发展提出了具体的措施，这一系列在国家层面上与数据要素相关的举措充分表现了中央对数据要素数乘效应的肯定。

在数字经济背景下，以大数据、人工智能等为代表的数字技术不断发展，使得消费模式发生巨大变革，数字化消费和支付方式提高了人们的消费意愿，成为解决消费问题、扩大经济需求的突破点，数据要素也成为推动新消费发展的关键。同时，网络的普及和数据的发展也促进了农村消费潜力的释放，促进农村消费结构升级，从而使城乡居民消费差距不断缩小。为此，本文主要研究数据要素的发展水平及它如何作用于城乡居民的消费差距。以下是本文的研究框架。

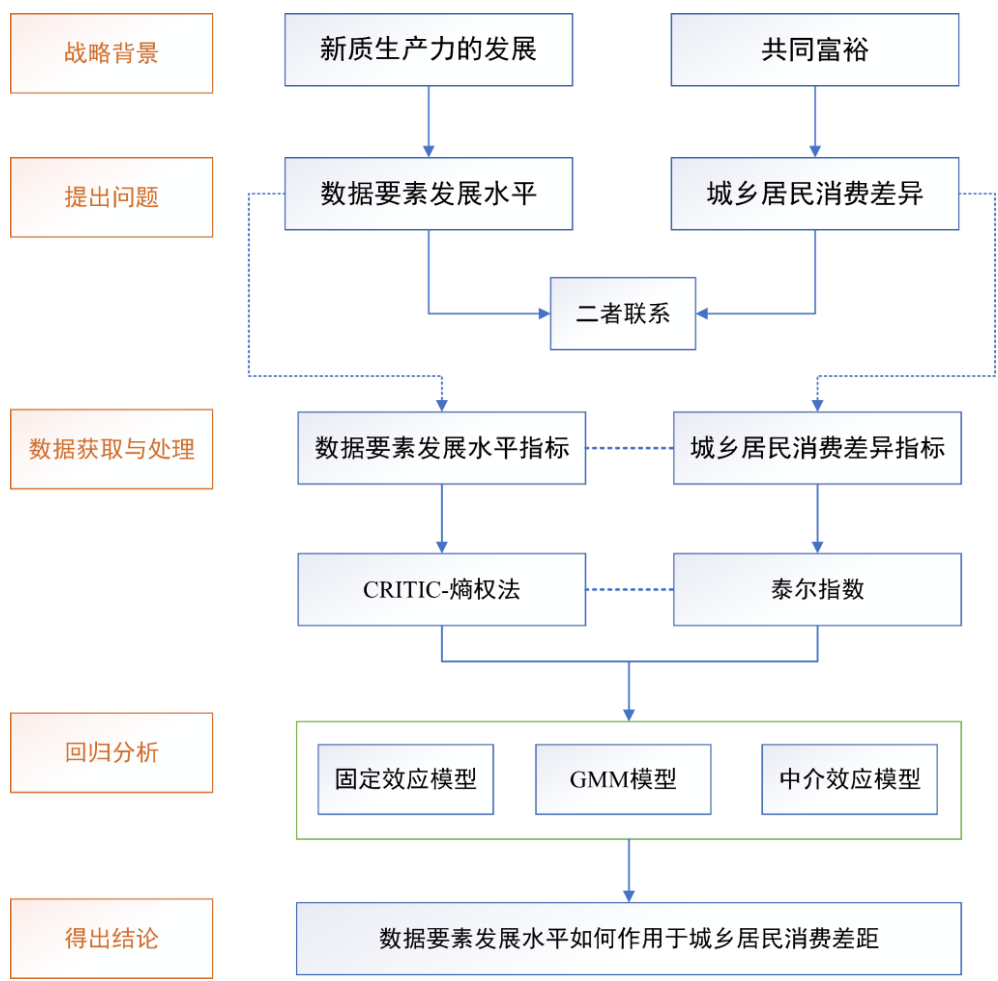

图 1 研究框架

## 二、文献综述

### （一）数据要素

潘宏亮、赵兰香、叶璐（2024）研究了我国数据要发展水平的测度及时空演进，表明我国数据要素发展水平总体上呈增长趋势，但区域差异呈现"东—中—西"一次递减的格局。徐翔、田晓轩、历克奥博、陈斌开（2024）通过改良现有估计方法对2012—2019年各省份数据要素投资进行测算。孙萍（2024）基于279个地级市面板数据的空间计量分析研究了数据要数发展对供给体系质量的影响。屈林发、严梓涵、罗健秋（2022）运用熵值法和CRITIC法研究长江经济带数据新动能统计测度及发展路径。方元欣、郭骁然（2020）基于市场角度，分析数据要素的经济价值和社会价值，探究了数据要素价值的影响因素，对现有的数据要素价值评估体系方法进行了前瞻性分析。当前，我国数据要素市场还处于起步阶段，迫切需要将数据资源优势转化为经济增长发展优势。

### （二）城乡居民消费差异

张莹、何旺、张卓群（2022）采用Dagum基尼系数和Kernel核密度估计对我国居民人均消费支出的区域差异及分布动态演进进行实证分析。章印、王永瑜（2023）基于消费环境、消费能力、消费信心和消费结构构建四个子系统构建中国居民消费潜力评价指标体系，并运用Dagum基尼系数和核密度估计方法揭示了居民消费潜力及各子系统的区域差异及分布动态演进规律。杨璐衣（2023）通过动态面板模型、门槛面板模型和空间面板模型探究数字金融普惠对城乡居民消费质量差距的影响。徐燕平（2024）构建了数据要素资源禀赋综合评价体系指数，研究了数据要素资源禀赋对城乡居民消费潜力的影响效应。缪言、曾晶、白仲林（2023）将数据要素成本、数字化技术与数据质量外生冲击引入包含互联网平台主体的DSGE模型，模拟分析消费、投资、可支配收入和居民消费倾向等变量的演变路径。

### （三）对现有文献的评论

从已有研究来看，国外聚焦数据要素与城乡居民消费差距关系的相关研究主要关注以下几个方面：一是城乡居民消费水平对参与数据要素发展的影响，以及

经济发展等因素对城乡互联网使用程度与普及力度的影响，在这个层面较多关注城乡数字鸿沟问题（Chinn M D 等，2010）；二是数据要素发展对微观个人的增收效应及区域发展影响（Paul D B B 等，2008）；三是数字技术对收入分配的影响，相关研究集中在发达国家，且未达成一致认识（Bauer J M 等，2018）。国内研究在探讨城乡居民消费差距影响因素上，一致认为城镇化、金融发展、人口结构、交通基础设施等是主要影响因素（吴鹏等，2017）。

可以发现，现有相关研究对数据要素这一新变量的关注度不足，多数从互联网技术发展与运用层面研究对收入分配的影响，但结论不一。当前较少有学者将数据要素发展水平与城乡居民消费差异结合研究，本文分析数据要素发展水平对城乡居民消费差距的影响，希望这些结论可以为下一阶段减小城乡居民消费差异、实现共同富裕提供帮助。同时，由于学术界尚未对数据要素发展水平测度达成一致，现有研究难以精准反应数据要素发展，因此对于测度的研究依旧有改进和完善的空间。

## 三、数据来源与描述

### （一）数据来源

本文从我国的实际情况出发，结合数据可得性，选取了 2013-2020 年我国 30 个省份（不包括港澳台、西藏）的相关数据。在样本数据来源上，本文所涉及的数据均为国家统计局公布的《中国地区投入产出表》《中国统计年鉴》《中国经济普查年鉴》中的数据，并利用线性插值法补齐缺失年份数据。由于人均 GDP 的数据间差距较大，为了消除异方差影响，采用对数化方法对变量进行处理。

在数据要素发展水平指标选取方面，本文参照行业通用指标，根据多位学者的研究成果制定了数据要素发展水平指标。这一指标主要研究了数据基础支撑、数据转化能力、数据行业应用三个维度，较为全面的反映了我国各省市的数据要素发展水平。具体指标体系如下表所示。

**表 1 数据要素发展水平指标体系**

| 维度指标 | 基础指标 | 指标说明 | 指标权重 |
| --- | --- | --- | --- |
| 总指数 | 衡量各个省份的数据要素发展水平 | | |
| 数据基础支撑 | 域名数 | 网络连接 | 0.0638 |
| | IPV4 地址数 | 网络基础 | 0.0755 |
| | 互联网接入端口数 | 通信能力 | 0.337 |
| | 单位面积光缆长度 | 通信基础 | 0.0747 |
| | 每百家企业拥有网站数 | 企业基础信息 | 0.0137 |
| 数据转化能力 | 技术市场成交额 | 数据创新能力 | 0.0811 |
| | 软件产品收入 | 数据管理能力 | 0.0709 |
| | 信息技术服务收入 | 数据应用能力 | 0.0786 |
| | 电子信息制造业业务收入 | 数据产业产出 | 0.0945 |
| | 电子信息制造业企业数 | 数据产业规模 | 0.0978 |
| | 信息传输、软件和信息技术服务业从业人数 | 数据产业规模 | 0.0539 |
| | 人工智能企业数 | 数据创新能力 | 0.0861 |
| 数据行业应用 | 电子商务交易额 | 商贸领域应用 | 0.0579 |
| | 有电子商务交易活动企业比重 | 企业应用 | 0.0206 |
| | 数字金融覆盖广度指数 | 数字金融领域应用 | 0.0205 |
| | 数字金融覆盖深度指数 | 数字金融领域应用 | 0.0228 |
| | 数字金融数字化程度 | 数字金融领域应用 | 0.0222 |
| | 数字经济发展水平 | 数字经济领域应用 | 0.0315 |

## （二）数据要素发展水平结果及分析

将除西藏外其余中国 30 个省市 2013-2022 年的指标数据代入上述模型，得出各省市数据要素发展水平的综合评价结果，并依据相关数据绘制各省市数据要素发展水平三维立体柱状图，见下图。

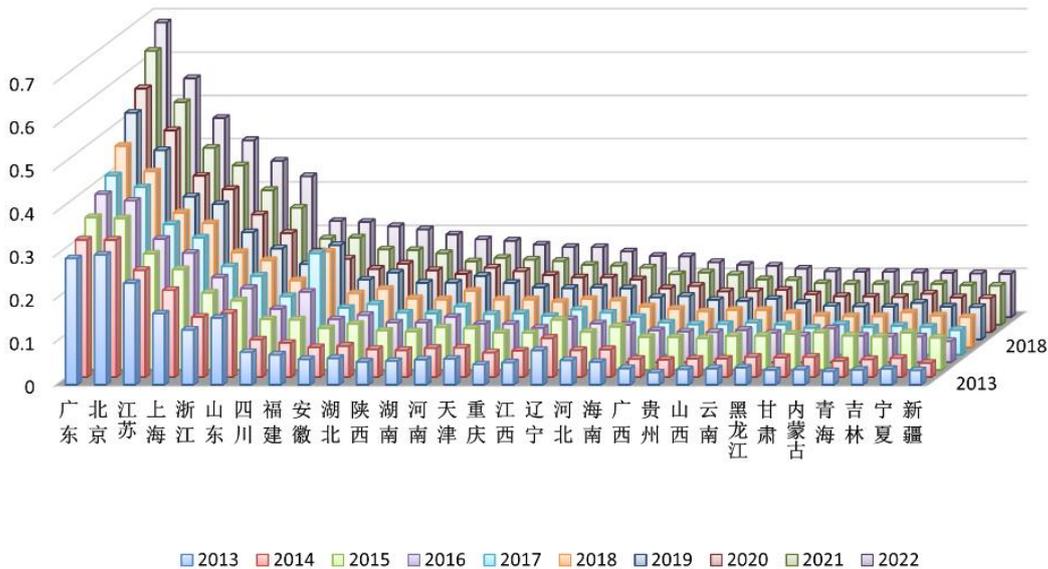

**图 2　我国 2013-2022 年各地区数据要素发展水平**

从历年各省数据要素发展水平来看，各省数据要素发展水平整体呈现稳步增长的趋势，但发展存在一定的不平衡性。西北和西南的大部分地区数据要素发展较为缓慢，且发展水平明显落后于经济发达地区。就省份而言，个别省份如福建省，数据要素发展呈现出先上升后下降再上升的趋势；广东、北京、江苏、上海、浙江、山东等地的历年发展水平较其他省市地区基本处于领先地位，且均超过了 0.4，这与长三角、珠三角、京津冀等实力雄厚的经济带密切相关。尤其是广东省与北京省，数据要素发展水平已经超过 0.5。而历年来江苏、上海、浙江三个省市数据要素发展水平排名保持不变，且浙江在 2015 年发展水平超过山东，这与长江三角洲区域一体化发展密切相关。到 2022 年，黑龙江、甘肃、内蒙古、青海、吉林、宁夏等地区数据要素发展水平仅超过 0.1，而新疆自治区数据要素发展水平只达到 0.0995，与高水平发展地区存在明显差异。

# 四、研究方法及模型建立

## （一）研究方法概述

### 1. 数据标准化处理

由于不同指标单位不同，为消除不同指标之间量纲的差异，增强指标之间的可比性，首先对所有数据进行标准化处理。具体步骤如下：

对于正向指标的处理选择：

$$x_{ij} = \frac{X_{ij} - X_{min}}{X_{max} - X_{min}} \tag{1}$$

对于负向指标的处理选择：

$$x_{ij} = \frac{X_{max} - X_{ij}}{X_{max} - X_{min}} \tag{2}$$

其中 $i = 1,2,\ldots,m$，$j = 1,2,\ldots n$，$X_{ij}$ 为各地区历年不同指标的原始数据，$X_{max}$ 为指标 $j$ 的最大值，$X_{min}$ 为指标 $j$ 的最小值，$x_{ij}$ 为标准化处理后的无量纲数据。

### 2. CRITIC—熵权法组合权重模型

CRITIC 赋权法是一种以对比强度和冲突性对比强度确定各个指标权重的客观赋权方法。其中，对比强度以各指标标准差的形式来表示。即标准差大小表明了同一指标的数据差异大小；冲突性是以各指标相关性为基础，如两个指标相关系数越大，则两个指标冲突性较低。

熵权法是一种基于信息熵概念的客观赋权法。该方法将信息熵作为指标的不确定性度量，认为在一组指标中，信息的熵值越小，说明该指标的离散程度越大，因此它在综合评价中的权重也就越大。

CRITIC—熵权法组合权重模型是一种结合 CRITIC 赋权法和熵权法的赋权方法，这样做既利用了数据的离散程度信息，又削弱了指标相关的影响，适用于多指标客观评价。具体步骤如下：

首先，根据 CRITIC 法计算各指标权重：

$$C_j = \frac{\sigma_j}{\bar{x}_j} \sum_{k=1}^{n}(1 - |r_{kj}|) \tag{3}$$

$$\omega_{1j} = \frac{c_j}{\sum_{j=1}^{n} c_j} \tag{4}$$

其中 $\sigma_j$ 为第 $j$ 个指标的标准差，$\bar{x}_j$ 为第 $j$ 个指标的平均值，$r_{kj}$ 为第 $k$ 个指标

与第$j$个的相关系数，$\omega_{1j}$为第$j$个指标的 CRITIC 权重。

接着，根据熵权法确定各指标权重：

$$P_{ij} = \frac{x_{ij}}{\sum_{i=}^{m} x_{ij}} \tag{5}$$

$$e_j = -\frac{1}{lnm}\sum_{i=1}^{m} P_{ij}lnP_{ij} \tag{6}$$

$$\omega_{2j} = \frac{1-e_j}{\sum_{j=1}^{n}(1-e_j)} \tag{7}$$

其中，$P_{ij}$表示第$i$个评价对象第$j$项指标所占比重，$e_j$表示第$j$项指标的信息熵，$\omega_{2j}$表示第$j$项指标的熵权法权重。

最后，根据 CRITIC 赋权法和熵权法所得第$j$项指标的综合权重为：

$$\omega_j = \beta\omega_{1j} + (1-\beta)\omega_{2j} \tag{8}$$

本文假设两种方法同样重要，即$\beta = 0.5$

## 3. 泰尔指数

泰尔指数是一种用来衡量收入或财富分配不平等程度的指数。它可以帮助我们了解一个群体中个体之间收入或财富的不平等程度。泰尔指数的数学定义如下：

给定一个包含$n$个个体的群体，其中第$i$个个体的收入为$y_i$，总收入为$y = \sum_{i=1}^{n} y_i$。泰尔指数的公式为：

$$T = \frac{1}{n}\sum_{i=1}^{n}(\frac{y_i}{Y}ln\frac{y_i}{Y}) \tag{9}$$

其中，ln 表示自然对数。泰尔指数的取值范围通常在 0 和 1 之间，值越接近 1 表示财富或收入分配越不平等。

泰尔指数的计算需要对个体收入进行加权平均，通过这种加权平均可以反映出不平等程度。

## 4. 固定效应模型

固定效应回归模型是一种面板数据分析方法。在面板数据线性回归模型中，如果对于不同的截面或不同的时间序列，只是模型的截距项是不同的，而模型的斜率系数是相同的，则称此模型为固定效应模型。其可以分为三类：个体固定效应模型、时点固定效应模型及时点个体固定效应模型。

$$y_{it} = \lambda_i + \sum_{k=2}^{k} \beta_k x_{k,i,t} + \mu_{i,t} \tag{10}$$

$$y_{it} = \gamma_t + \sum_{k=2}^{k} \beta_k x_{k,i,t} + \mu_{i,t} \tag{11}$$

$$y_{it} = \lambda_i + \gamma_t + \sum_{k=2}^{k} \beta_k x_{k,i,t} + \mu_{i,t} \tag{12}$$

## 5. 系统 GMM 回归模型

动态面板模型设定中将被解释变量的滞后项作为解释变量引入到回归模型中，使得模型具有动态解释能力，但模型中存在内生性问题。为了解决这一内生性，Arellano 和 Bond 提出了利用工具变量来推导相应矩条件的广义矩（GMM）方法，所谓的"差分 GMM 方法"。虽然差分 GMM 方法降低了内生性对模型估计带来的影响，但在有限样本条件下，差分 GMM 方法存在严重的"弱工具变量"问题。Arellano 和 Blundell 等人提出了更完美的"系统 GMM 方法"。

系统 GMM 方法对原水平模型和差分变换后的模型同时进行估计，系统 GMM 能够修正未观察到的个体异质性问题、遗漏变量偏差、测量误差和潜在的内生性问题，这些问题在使用混合 OLS 和固定效应方法时常常会影响模型的估计效果。系统 GMM 方法还能减少由于使用一阶差分 GMM 估计方法带来的潜在偏误和不精确性。

为了模型的准确性，必须通过第一个检验，确保矩条件不被过度约束，工具变量的个数不能超过内生变量的个数。需要模型的扰动项具有显著的一阶相关和不显著的二阶自相关。为此，在一阶差分残差中使用了一阶序列相关和二阶序列相关的 Arellano-Bond 检验，即 AR(1)小于 0.1，AR(2)大于 0.1。二是过度识别约束检验，该检验主要是判断系统 GMM 估计中所采用的工具变量是否整体有效，常采用 Sargan 检验或者 Hansen 检验进行判断，其原假设是所有的工具变量都是外生的。因此，若工具变量是有效的，则不应拒绝原假设。

## 6. 中介效应模型及 Sobel 检验

中介效应指 X 对 Y 的影响是通过 M 实现的，也就是说 M 是 X 的函数，Y 是 M 的函数（Y-M-X）。考虑自变量 X 对因变量 Y 的影响，如果 X 通过 M 影响变量 Y，则称 M 为中介变量。

Sobel 检验是中介效应检验方法的一种，当发现系数 a 或者系数 b 其中有一个不显著时，可以利用统计检验的方法检验系数乘积$ab$是否显著异于 0。Sobel 法就是通过构建系数乘积$ab$的统计量$z$来估计其置信区间，判断其是否显著异于 0。但是 Sobel 法构建的统计量的推导需要假设$\hat{a}\hat{b}$服从正态分布，假设要求较高。

$$z = \hat{a}\hat{b}/s_{ab} \tag{13}$$

$$s_{ab} = \sqrt{\hat{a}^2 s_b^2 + \hat{b}^2 s_a^2} \tag{14}$$

## （二）模型建立

**1. 被解释变量**

在现有的文献中，一共有三种方法测算城乡居民消费差距：一是用城乡人均消费支出占比来衡量；二是用基尼系数，它需要在不同阶层人群分解，它衡量的是整体的差距，对中间阶层的变动更敏感；三是用泰尔指数，可以衡量组内差距和组间差距对总差距的贡献。本文采用泰尔指数衡量各个省的城乡居民消费差距，测算公式如下：

$$GAP_{i,t} = \sum_{j=1}^{2} \frac{Y_{i,j,t}}{Y_{i,t}} \times \ln\left[\frac{Y_{i,j,t}}{Y_{i,t}} \div \frac{X_{i,j,t}}{X_{i,t}}\right] \tag{15}$$

其中，$GAP_{i,t}$表示某省 $i$ 第 $t$ 年的消费差距泰尔指数，$j = 1,2$ 分别代表城镇与乡村，$Y_{i,j,t}$表示省份 $i$ 在 $t$ 年城镇（乡村）居民的消费支出，$Y_{i,t}$表示该省第 $t$ 年总的消费支出。$X_{i,j,t}$表示省份 $i$ 在 $t$ 年城镇（乡村）的人口数，$X_{i,t}$表示该省在第 $t$ 年的总人口数。测算得到的结果可以衡量某省份 $i$ 在第 $t$ 年的城乡居民消费差距。泰尔指数的值在 0 到 1 之间，数值越大，则说明差距越大。

**2. 解释变量**

本文的目的是为了探究数据要素对城乡居民消费差距的影响，因此解释变量应能衡量各个省的数据要素发展水平，即数据要素发展水平。本文也在第三章对数据要素发展水平进行了说明及测度。

**3. 中介变量与控制变量**

本文的中介变量为城乡收入差距。参考上文城乡消费差距，本文选取了《中国统计年鉴上》上相关的收入和人口情况的数据，采用泰尔指数测算城乡居民收入差距。

为了确保结果的准确性，结合实际情况，本文参考了前人对城乡居民消费差距的影响因素研究（李学凯，2022；周莹，2022；方正之，2023；高雪，2024），采用财政支出（Fis），人口结构（Tdr），对外开放水平（Open），人均 GDP（Pgdp），受教育程度（Edu）这五个量为控制变量。

## 五、回归结果分析

### （一）固定效应模型

本文选取具有固定效应的面板模型。因此，为验证城乡居民消费差异与各个控制变量之间的关系，本文构建了以下模型：

$$C_{i,t} = \beta_1 \text{Dig}_{i,t} + \beta_2 \text{Fis}_{i,t} + \beta_3 \text{Tdr}_{i,t} + \beta_4 \text{Open}_{i,t} + \beta_5 \text{Pgdp}_{i,t} + \beta_6 \text{Edu}_{i,t} + u_i + \varepsilon_{i,t} \quad (16)$$

在模型中，C 表示被解释变量城乡居民消费差异，$\text{Dig}_{i,t}$ 表示省份 $i$ 在第 $t$ 年的数据要素发展水平，Fis 表示财政支出，Tdr 表示人口结构，Open 表示对外开放水平，Pgdp 表示人均 GDP，Edu 表示受教育程度。根据上述建立固定效应面板回归模型，利用 Stata 回归得到如下结果。

表 2  固定效应回归结果

| 变量 | （1） | （2） |
| --- | --- | --- |
| Dig | -0.2112*** | -0.1171*** |
|  | (0.0000) | (0.0249) |
| Fis |  | 0.2434* |
|  |  | (0.1295) |
| Tdr |  | -0.0011*** |
|  |  | (0.0003) |
| Open |  | -0.0437** |
|  |  | (0.0170) |
| Pgdp |  | 0.0640 |
|  |  | (0.1437) |
| Edu |  | -0.0118*** |
|  |  | (0.0021) |
| cons | 0.0802*** | 0.2003*** |
|  | (0.0000) | (0.0289) |
| $N$ | 300 | 300 |
| adj. $R^2$ | 0.4378 | 0.7094 |

注：***$p$<0.01，**$p$<0.05，*$p$<0.1，括号内为稳健标准误。

通过固定效应回归模型的分析，整体上看，数据要素发展水平与城乡居民消

费差异呈现显著负相关性。根据第（2）列的结果可看出，数据要素每发展1单位，城乡居民消费差异会缩小0.1171个单位，表明数据要素发展水平促进城乡居民消费差异的减小。其中人口结构、对外开放水平、受教育程度等控制变量对减小城乡居民消费差异呈现出很好的显著性。

## （二）稳健性检验

稳健性检验包括替换变量法、剔除部分样本、缩尾处理等多种方法。本文选择替换被解释变量进行稳健性检验分析。通过查阅文献可知，目前测算城乡消费差异的方法主要为泰尔指数、基尼系数以及城乡居民消费差异（城镇与农村消费支出比例）。选取城乡居民消费差异为被解释变量做基准回归，结果如下表所示。

表3 稳健性检验结果

| 变量 | （2） |
| --- | --- |
| Dig | -0.3673* |
|  | (0.2067) |
| Fis | 1.1381* |
|  | (0.6362) |
| Tdr | -0.0046*** |
|  | (0.0016) |
| Open | -0.2362*** |
|  | (0.0821) |
| Pgdp | -0.0772 |
|  | (0.8808) |
| Edu | -0.0645*** |
|  | (0.01042) |
| cons | 0.9519*** |
|  | (0.1339) |
| $N$ | 300 |
| adj. $R^2$ | 0.6138 |

注：***$p<0.01$, **$p<0.05$, *$p<0.1$，括号内为稳健标准误。

由上述结果可知，核心解释变量数据要素发展水平回归系数为-0.3672，且通过显著性检验，同时其余控制变量也在不同水平通过显著性检验。检验结果表明，

即使替换被解释变量测算方法，回归结果依旧符合预期，即数据要素发展对缩小城乡消费差异存在促进作用。

## （三）系统 GMM 回归模型

根据杜森贝里提出的相对收入消费理论，消费者的消费支出不仅仅依赖于他的现期收入，还依赖于其他人的收入水平和他过去曾达到的最高收入水平。因此，我们引入被解释变量城乡消费差异的滞后一期，建立如下模型进行回归分析：

$$C_{i,t} = \alpha + \rho_0 C_{i,t-1} + \rho_1 Dig_{i,t} + \rho_2 Fis_{i,t} + \rho_3 Tdr_{i,t} + \rho_4 Open_{i,t} + \rho_5 Pgdp_{i,t} + \rho_6 Edu_{i,t} + u_i + \varepsilon_{i,t} \tag{17}$$

其中 $C_{i,t-1}$ 表示城乡居民消费差异的最后一期。回归结果如下：

表 4  系统 GMM 模型回归结果

| 变量 | （1） |
| --- | --- |
| L.C | 0.642*** |
|  | (0.0772) |
| Dig | -0.0196*** |
|  | (0.0072) |
| Fis | 0.0986* |
|  | (0.0579) |
| Tdr | -0.0005*** |
|  | (0.0002) |
| Open | -0.0101** |
|  | (0.0052) |
| Pgdp | 0.0257 |
|  | (0.0361) |
| Edu | -0.0040*** |
|  | (0.0014) |
| cons | 0.0670*** |
|  | (0.0190) |
| AR(1) | -3.49*** |
| AR(2) | -0.51 |
| Hansen | 21.74 |
| Observations | 270 |

注：***$p<0.01$, **$p<0.05$, *$p<0.1$，括号内为稳健标准误。

由结果可知，AR(1)通过显著性检验，AR(2)没有通过显著性检验，表明模型拒绝不存在一阶自相关的原假设，接受不存在二阶自相关的假设，因此模型较好地克服了内生性问题。同时，结合 Hansen 检验结果，GMM 模型选取的工具变量有效，不存在过度识别的问题。

此外，数据要素发展水平在此模型中对城乡居民消费差异的回归系数显著为负，这与进行基准回归时得出的结论相一致，即当数据要素发展水平提高 1 个单位时，城乡居民消费差异将会减少 0.0196 个单位，因此数据要素的发展对缩小城乡消费差异存在促进作用。而消费差异滞后一期系数显著为正，可见当期消费差异明显受滞后一期消费差异的影响，进一步证明建立系统 GMM 动态回归模型的合理性。

## （四）中介效应模型

为了进一步探究数据要素是否会通过城乡居民收入差距来影响城乡居民之间的消费差距，我们构建检验中介效应检验模型。中介效应是指变量间的影响关系不是直接的因果链，而是通过变量 M 间接影响产生，变量 M 称为中介变量。中介效应的检验步骤如下图所示。

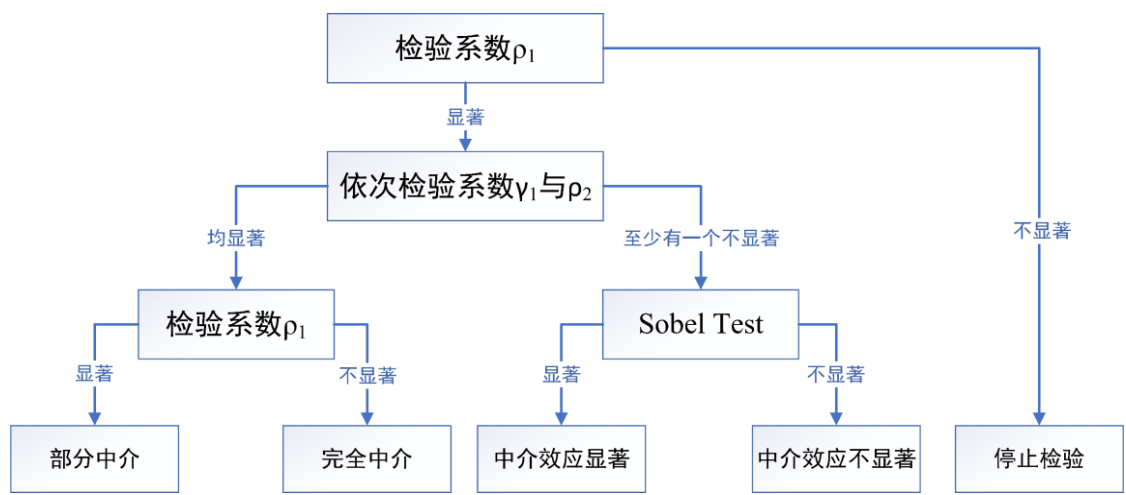

**图 3　中介效应检验步骤**

随后本文参考温忠麟等相关研究构建如下模型进行检验：

$$C_{i,t} = \alpha_0 + \alpha_1 Dig_{i,t} + \sum \beta_j X_{j,i,t} + \mu_i + \varepsilon_{i,t} \tag{18}$$

$$I_{i,t} = \gamma_0 + \gamma_1 Dig_{i,t} + \sum \beta_j X_{j,i,t} + \mu_i + \varepsilon_{i,t} \tag{19}$$

$$C_{i,t} = \rho_0 + \rho_1 Dig_{i,t} + \rho_2 I_{i,t} + \sum \eta_j X_{j,i,t} + \mu_i + \varepsilon_{i,t} \tag{20}$$

其中，下标 $i$ 和 $t$ 分别表示省份与时间；变量 $C_{i,t}$ 表示城乡居民消费差异；变量 $I_i$ 表示城乡居民收入差距；变量 $Dig_{i,t}$ 表示数据要素发展水平；$\varepsilon_{i,t}$ 表示随机误差项。中介效应检验路径如图 3 所示，根据中介效应检验流程，首先对公式（17）进行基准回归，探究数据要素发展水平对城乡居民消费差距的影响，然后对公式（18）进行回归，检验数据要素发展水平与中介变量城乡居民收入差距之间的关系，若系数为负，则数据要素会缩小城乡居民收入差距；最后，再对公式（19）进行回归分析，同时探究数据要素发展水平、城乡居民收入差距对城乡居民消费差距的影响，若 $\gamma_1 < 0$，$\alpha_1 < 0$，$\rho_1 < 0$ 且各项系数均显著，则说明数据要素通过缩小城乡居民收入差距影响城乡居民消费差距的机制是存在的。回归结果如下表所示。

表 5 中介效应回归结果

|  | 模型(1) 城乡消费差距(C) | 模型(2) 城乡收入差距(I) | 模型(3) 城乡消费差距(C) |
| --- | --- | --- | --- |
| Dig | -0.1171*** (0.0001) | -0.0553** (0.0449) | -0.0612*** (0.0029) |
| Fis | 0.2434* (0.0849) | 0.1871* (0.0863) | 0.0543 (0.3285) |
| Tdr | -0.1065*** (0.0038) | -0.0862*** (0.0029) | -0.0194 (0.3425) |
| Open | -0.0437** (0.0213) | -0.0435*** (0.0018) | 0.0002 (0.9780) |
| Pgdp | 0.0641 (0.6751) | -0.1056 (0.5043) | 0.1708** (0.0126) |
| Edu | -0.0118*** (0.0000) | -0.0107*** (0.0000) | -0.0011 (0.3015) |
| I |  |  | 1.0106*** (0.0000) |

|  | 模型(1) | 模型(2) | 模型(3) |
|---|---|---|---|
|  | 城乡消费差距(C) | 城乡收入差距(I) | 城乡消费差距(C) |
| cons | 0.2004*** | 0.2194*** | -0.0214 |
|  | (0.0000) | (0.0000) | (0.2910) |
| *N* | | 300 | |
| *adj. R²* | | 0.9550 | |
| Sobel 检验 | 0.8783 | 0.9657 | -3.083*** |
|  |  |  | （0.002） |

注：***$p<0.01$, **$p<0.05$, *$p<0.1$，括号内为稳健标准误；Sobel检验报告z值，括号内为p值。

由上述回归结果可以看出，模型（1）的回归系数为-0.1171，显著为负；模型（2）的回归系数为-0.0553，显著为负。这说明了数据要素的发展在1%水平上显著缓解了城乡居民消费差距的扩大。模型（3）中，加入城乡收入差距进行回归，收入差距的系数为1.0106，通过了1%水平上的显著性检验，则说明了城乡居民收入差距与消费差距有显著的正向关系。同时数据要素的回归系数在10%的水平下显著为负，且相较于基准回归的结果，纳入城乡收入差距变量后，数据要素对城乡消费差距的缩小程度有所减少。由此可以得出：城乡收入差距在数据要素发展水平与城乡消费差距之间的中介效应存在。由计算可得：数据要素对城乡消费差距的总效应为-0.1171，中介效应为-0.0559，中介效应占总效应的47.75%。

因为逐步回归可能会漏掉一部分中介效应，故为使结论更加可靠，我们同时进行了Sobel检验，检验通过了1%水平的显著性检验，再次说明了收入差距这个中介作用的存在，这也是数据要素缩小城乡消费差距的一个重要途经。

## 六、结论与建议

### （一）结论

**1. 数据要素发展水平整体呈现上升趋势，但东、中、西三大区域的数据要素发展存在差异，且有两级分化的发展态势**

本文通过构建数据要素发展水平评价指标体系，分别从数据投入基础、数据转化能力以及数据应用能力三个方面考虑，运用CRITIC—熵权法赋权得出30个省市的数据要素发展水平，即数据要素发展水平整体呈现上升趋势。数据要素发达地区集中在经济发达的省市，东部地区数据要素发展水平较高，中部地区和西部地区次之，东北部地区数据要素发展水平缓慢，全国呈现发展不均衡局面。

**2. 数据要素的发展能够显著促进缩小城乡居民消费差距，且存在中介效应**

通过固定效应回归模型的分析，整体上看，数据要素发展水平与城乡居民消费差异呈现显著负相关性，表明数据要素发展水平促进城乡居民消费差异的减小。且通过中介效应结果可知，城乡收入差距在数据要素与城乡消费差距之间的中介效应存在，且收入的中介效应达到 47.75%。因此，数据要素的发展可以通过促进缩小城乡收入差异，进而促进缩小城乡居民消费差异。

**3. 城乡居民消费差异滞后一期显著影响当期消费差异**

通过系统 GMM 模型回归分析结果可以得知，在引入城乡消费差异滞后一期进行回归后，回归系数显著为正，且数据要素发展水平回归系数同固定效应与中介效应回归系数一致，并在 1%的显著性水平上显著，呈现出良好的稳健性。

### （二）建议

针对上述结论，本文提出以下建议：

**1. 推动数据要素发展**。数字技术是推动数据要素发展的关键驱动力。包括数据采集、存储、处理、分析和可视化等各个环节的技术创新，尤其是人工智能、机器学习、大数据分析等领域的进步，能够极大地促进数据要素的发展。因此，各地方政府应鼓励引导发展数字技术，通过政策推动各地区数据要素基础设施的建设，以激发创造数据要素价值的潜力与活力，发挥数据要素这一重要生产力。

**2. 完善数据要素区域协调发展机制**。国家应大力推动中西部区域数据要素

发展，促进跨区域数据基础设施的建设和项目合作。同时应注重城乡之间基础设施差距，大力发展数据基础设施在农村中的作用以弥补数字鸿沟。

**3. 通过数据要素市场促进城乡收入分配公平。** 通过数据资源的有效利用和市场化交易，实现收入在不同参与者之间的公平分配。可以建立公开、透明的数据市场机制，使得数据要素在分配中全面发挥作用，通过减少信息差以缩小城乡居民收入差异。

**4. 发展数字经济与电子商务。** 加强数字经济和电子商务在城乡地区的发展，提升城乡居民的消费便利性和体验。通过数据要素，推动电子商务在农村地区的发展，拓展农村居民消费选择范围。

**5. 促进城乡融合发展。** 完善城乡公共服务体系，加大对农村教育、医疗、文化等公共服务的投入，缩小城乡公共服务差距，推动农村产业与城市产业融合发展，实现产业互补共赢。

# 参考文献